\documentclass[aps,pre,showpacs,showkeys,twocolumn,groupedaddress]{revtex4-1}
\usepackage{amssymb}
\usepackage{graphicx}
\usepackage{amsmath}
\usepackage{setspace}

\begin{document}


\title{Size-Dependency of Income Distributions and Its Implications}


\author{Jiang Zhang}
\email{zhangjiang@bnu.edu.cn}
\homepage[]{http://www.swarmagents.cn/jake}
\author{You-Gui Wang}
\email{ygwang@bnu.edu.cn}
\affiliation{Department of Systems Science, School of Management,
Beijing Normal University}


\date{\today}

\begin{abstract}
This paper highlights the size-dependency of income distributions,
i.e. the income distribution versus the population of a country
systematically. By using the generalized Lotka-Volterra model to fit
the empirical income data in the United States during 1996-2007, we
found an important parameter $\lambda$ can scale with a $\beta$
power of the size (population) of U.S. in that year. We pointed out
that the size-dependency of the income distributions, which is a
very important property but seldom addressed by previous studies,
has two non-trivial implications: (1) the allometric growth pattern,
i.e. the power law relationship between population and GDP in
different years, which can be mathematically derived from the
size-dependent income distributions and also supported by the
empirical data; (2) the connection with the anomalous scaling for
the probability density function in critical phenomena since the
re-scaled form of the income distributions has the exactly same
mathematical expression for the limit distribution of the sum of
many correlated random variables asymptotically.
\end{abstract}

\pacs{89.65.Gh,89.75.Da,89.75.Kd}
\keywords{Income Distribution, Size-Dependency, Allometric Growth,
Anomalous Scaling}

\maketitle

\section{Introduction}
\label{sec.introduction}

The power law distribution of incomes in a nation is one of the most
important universal patterns found in economic systems due to the
seminal work of Pareto\citep{pareto_cours_1964}. It is suitable for
not only incomes and wealth in different countries and different
years
\citep{yakovenko_colloquium:_2009,clementi_power_2005,ding_power-law_2007,silva_temporal_2004}
but also other complex systems e.g.
languages\citep{zipf_selected_1932} and complex
networks\citep{albert_statistical_2002}. Although this statistical
law is supported by many empirical data\citep{souma_universal_2000}
and theoretical
works\citep{richmond_review_2006,chatterjee_pareto_2004}, it can
only describe the distribution in high incomes. Some recent studies
have shown that the distribution for the great majority of
population can be described by an exponential function which is very
different from the power law in the high
incomes\citep{dragulescu_a._exponential_2001,silva_temporal_2004}.
Silva and Yakovenko \citep{silva_temporal_2004} defined these
different income intervals as thermal and super-thermal regions
whose dynamics may follow very different rules.

A recent paper of our group discussed how the income distribution
curves in China change with time\citep{xu_evidence_2010}, so a
problem arise that does the distribution curves change with the
system size? As we know, some early studies in family names have
pointed out that the distributions can change with the size of the
system\citep{miyazima_power-law_2000,kim_distribution_2005,anh_korean_2007}.
This size-dependency of distributions is also found in
languages\citep{bernhardsson_meta_2009}. In this paper, we try to
propose that the income distributions also have this size-dependency
property which means that the distribution curves change with the
system size (the population) systematically. In Section
\ref{sec.incomedistribution}, we used the revised form of the
generalized Lotka Volterra model to fit the empirical income data of
the United States during 1996-2007. In this formula, we inserted a
scale factor $\lambda$ which changes with the population as a power
law with exponent $\beta$ in different years. So the size-dependency
of income distribution is implicit revealed by this power law
relationship.

In Section \ref{sec.fromincometoallometry}, we also pointed out that
the size-dependency of income distributions actually implies the
power law relationship between population and GDP, i.e. the
allometric growth(scaling) phenomenon which is also found in various
complex systems such as ecological
systems\citep{nordbeck_urban_1971,kleiber_body_1932,brown_scaling_2000,west_revieworigin_2005},
cities\citep{isalgue_scaling_2007,bettencourt_invention_2007,bettencourt_growth_2007}
and
countries\citep{roehner_macroeconomic_1984,zhang_allometric_2010}.
And we have also tested this relationship by the empirical data.
Some studies in family names\citep{miyazima_power-law_2000} and
languages\citep{lu_zipfs_2010-1,leijenhorst_formal_2005} have linked
the patterns of power law distributions and power law relationships
of two variables. However, in this study, we argued that the
exponent of the power law relationship between population and GDP
doesn't depend on the Pareto exponent of income distribution but the
size-dependency exponent.

Furthermore, the re-scaled form of income distribution curves can be
re-expressed as a generalized mathematical form in Section
\ref{sec.generalizedform}. This formula actually has been found to
describe the anomalous scaling of probability density function in
critical phenomena, e.g. spin
systems\citep{baldovin_central_2007,stella_anomalous_2010}, where
the re-scaled distribution form can be treated as a limit
distribution of the sum of a large number of correlated random
variables\citep{baldovin_central_2007,stella_anomalous_2010}.
Therefore, the size-dependency of income distribution also implies
the connection with the central limit theorem of correlated
variables.

\section{Size-Dependent Income Distributions}
\label{sec.incomedistribution}

\begin{figure}
\centerline {\includegraphics[width=9cm]{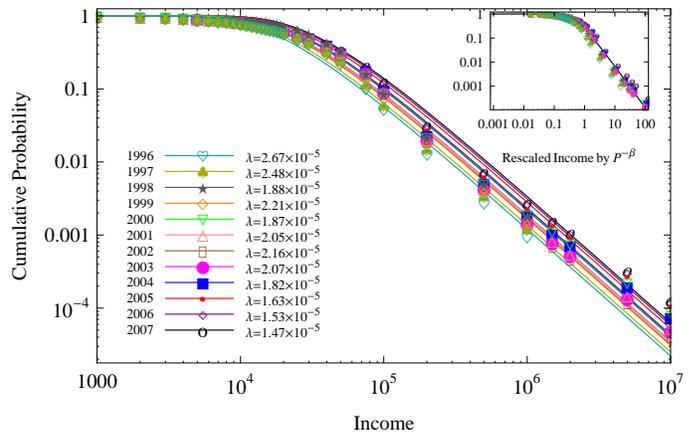}}
\vskip3mm \caption{Income Distributions of U.S. in the period of
1996-2007.}\label{fig.incomedistributions}
\end{figure}

The personal-income distribution data in the United States during
1996-2007 is available. This data is compiled by the Internal
Revenue Service (IRS) from the tax returns in the USA for the period
1996-2007(presently the latest available year \citep{url}). The
original data gives the percentage in given income intervals. We can
plot the cumulative distributions in Figure
\ref{fig.incomedistributions}. Notice that, the income data here is
just the nominal income so the inflation ingredients are not
excluded. Therefore, the GDP data we will use in the following parts
is also nominal or unadjusted for the effects of inflation.

\begin{figure}
\centerline {\includegraphics[width=8cm]{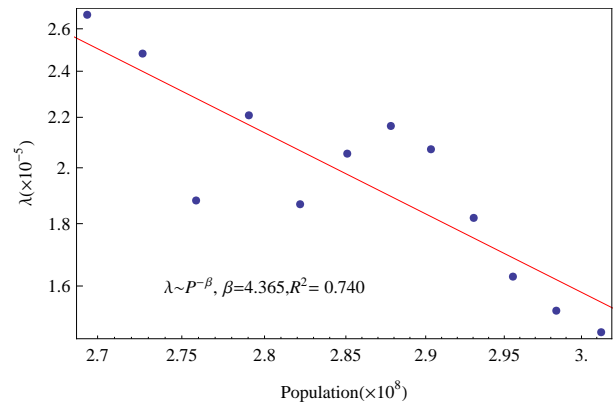}}
\vskip3mm \caption{The Power Law Relationship between the Population
and $\lambda$ during 1996-2007}\label{fig.lambdapopulation}
\end{figure}

As pointed out by \citep{silva_temporal_2004}, the distribution
curves exhibit exponential form in low incomes and power law
distribution in high incomes. However, we use the generalized
Lotka-Volterra
model\citep{malcai_theoretical_2002,richmond_review_2006} instead of
the method in \citep{silva_temporal_2004} to fit these data since
the generalized LV model only needs two free parameters. We assume
the density curves of income distributions in different years follow
the equation,
\begin{equation}\label{distribution}
f(x)=\lambda \frac{(\alpha -1)^{\alpha}}{\Gamma
(\alpha)}\frac{\exp(-\frac{\alpha -1}{\lambda x})}{(\lambda
x)^{1+\alpha}},
\end{equation}
where, $\alpha$ and $\lambda$ are parameters needed to be estimated.
$\Gamma()$ is the Euler gamma function. Note that, in the original
form of generalized-LV model\citep{malcai_theoretical_2002}, there
is no factor $\lambda$ since the main purpose of that paper is to
give an explanation of the shape of the income distribution.
However, we must insert this factor in Equation \ref{distribution}
because we care not only the shape of the income distribution curves
but also its dependency on size (the population of a country) in
different years. And this size-dependency property can only be
reflected by $\lambda$. In addition, $\alpha$ is the Pareto's
exponent in high incomes regime because Equation \ref{distribution}
has a truncated power law form. Nevertheless, we will use the
cumulative distribution function instead of Equation
\ref{distribution} directly to reduce the estimated errors.
\begin{equation}\label{cumdistribution}
C(x)=\int_{x}^{+\infty}f(x)dx=1-\frac{\Gamma(\alpha,\frac{\alpha-1}{\lambda
x})}{\Gamma(\alpha)},
\end{equation}
where the function $C(x)$ is the probability that a person whose
income is larger than $x$. Two steps fitting method is used in this
paper. At first, we apply Equation \ref{cumdistribution} to the
empirical data year by year and obtain the best estimations of
parameters $\alpha$ and $\lambda$. Here, the ``best'' means the
total distances between empirical data and theoretical curves on the
log-log coordinate are minimized. The $\alpha$s derived by the first
step are (1.59043, 1.60112, 1.61717, 1.67562, 1.75147, 1.76187,
1.71051, 1.61152, 1.80999, 1.95683, 1.87374, 1.92885), they
fluctuate around the mean value 1.74076. Second, we fix
$\alpha=1.74076$ and use the same method to obtain the best
estimation of $\lambda$ again. We will show that the size-dependency
and implications actually are independent on $\alpha$. The reason of
using this two steps fitting method is to get a better estimation of
$\lambda$ which is more important than $\alpha$.
From Figure \ref{fig.incomedistributions}, we can see that the
distributions change over time regularly. As time goes by, the
distribution curves shift. This trend is more obvious in the scaling
regions (high income tails). The relationship between the best
estimations of $\lambda$ and years is shown in the legend of Figure
\ref{fig.incomedistributions}.Furthermore, we know that the
population of U.S. increases with time in these years. Therefore,
$\lambda$ actually is the function of population at a given year.
This functional relationship can be presented by a power law
relationship between population and $\lambda$ as Figure
\ref{fig.lambdapopulation} shows.
From this figure, we observe an apparent trend that $\lambda$
decreases with population. This trend can be approximated by a power
law relationship between $\lambda$ and population.
\begin{equation}\label{eqn.lambdap}
\lambda \sim P^{-\beta},
\end{equation}
where $\beta$ estimated as $4.365$ is the slope of the line in
Figure \ref{fig.lambdapopulation}. Therefore, we conclude that the
income distributions are size(population) dependent. This dependency
is described by the power law relationship between the scale
parameter $\lambda$ and the population.

As a result, the income distributions in different years can be
re-scaled by $P^{-\beta}$,
\begin{equation}\label{eqn.rescaleddistribution}
f(x)\sim (P^{-\beta}) \frac{(\alpha -1)^{\alpha}}{\Gamma
(\alpha)}\frac{\exp(-\frac{\alpha -1}{P^{-\beta} x})}{(P^{-\beta}
x)^{1+\alpha}}.
\end{equation}
The re-scaled curve of income distribution is shown in the inset of
Figure \ref{fig.lambdapopulation}.

Although the power law relationship equation \ref{eqn.lambdap} is
acceptable because its $R^2$ is big enough, we know there are still
large deviations from the empirical data in Figure
\ref{fig.lambdapopulation}. We guess the main errors are from the
income distribution fittings. In Figure
\ref{fig.incomedistributions}, we know that there are some
deviations in the theoretical income distributions from the
empirical data. And these errors can of course influence the
estimations of $\lambda$s very dramatically since $\lambda$s are
very small. The second reason is we have very few samples here(only
11 years), so the noise in the original data can not be eliminated.
Thus, equation \ref{eqn.lambdap} is just an approximation, however
it can not prevent us to get an asymptotic theory. Next, we will
discuss the two important implications of this size-dependency.

\section{Power Law Relationship between Population and GDP}
\label{sec.fromincometoallometry}

We will show that the size-dependency of income distributions
implies the power law relationship between population and GDP. At
first, we know that the GDP of a country is proportional to the
total incomes of all people\citep{blanchard_macroeconomics_2000}.
Second, the total incomes can be read from the income distribution
curve. We can write down the equation,
\begin{equation}\label{eqn.gdp}
X \sim P\langle I\rangle,
\end{equation}
where, $X$ is the GDP, $P$ is the population of a given year. $I$ is
the random variable income in a given year. And $\langle I\rangle$
stands for the ensemble mean value of incomes. Therefore, $P\langle
I\rangle$ is just the total incomes of the whole country in the
given year.

Then we can calculate the mean income from the cumulative
probability function (Equation \ref{cumdistribution}) as follow,
\begin{equation}\label{eqn.meanincome}
\begin{aligned}
\langle I\rangle&=
\int_0^{+\infty}{xf(x)dx}=\int_0^{+\infty}{C(x)dx}=\frac{1}{\lambda}.
\end{aligned}
\end{equation}
Therefore,
\begin{equation}\label{eqn.gdplambda}
X \sim P/\lambda.
\end{equation}
Substituting Equation \ref{eqn.lambdap} into \ref{eqn.gdplambda}, we
get:
\begin{equation}\label{eqn.allometry}
X \sim P^{1+\beta}.
\end{equation}
Equation \ref{eqn.allometry} is just the power law relationship
(allometric growth) between population and GDP with an exponent
$1+\beta$. We have estimated the exponent $\beta \sim 4.365$ from
the income distributions. Therefore, we predict that the GDP is a
$5.365$ power of the population in the United States during
1996-2007.

\begin{figure}
\centerline {\includegraphics[width=8cm]{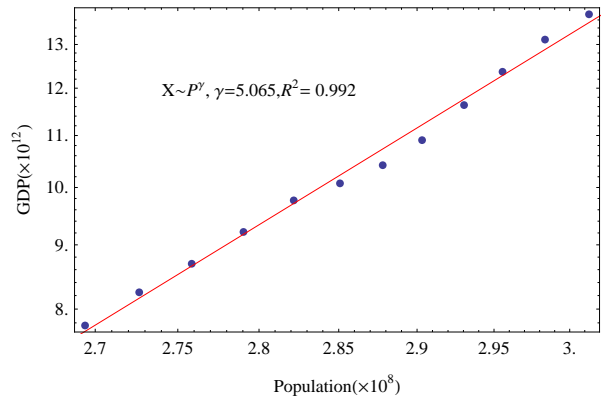}}
\vskip3mm \caption{Power Law Relationship between the Population and
the GDP of U.S. during 1996-2007}\label{fig.popgdptime}
\end{figure}

On the other hand, we can obtain the real data of the population and
the GDP of the United States during the given period. The two
variables have a power law relationship which is shown in Figure
\ref{fig.popgdptime}.
From the empirical data, we can estimate the power law exponent is
about 5.065 which is closed to the exponent we have predicted from
size-dependent income distribution data (the relative error is
$|5.365-5.065|/5.065
 \approx 6\%$). However, there is still a little deviation between
the empirical exponent and predicted one. The possible error sources
may include (1) the income curves; (2) the estimation of $\beta$;
(3) the deviation between GDP and total incomes.

\section{Generalized Size-Dependent Income Distribution}
\label{sec.generalizedform} One of an interesting fact which
deserves more attention is the size-dependency of income
distribution and its implication of the power law relationship
between population and GDP are independent on the Pareto exponent
$\alpha$ in the income distribution formula Equation
\ref{distribution}. Therefore, we can further hypothesize that the
size-dependency of distribution is a unique property independent on
the concrete form of the density function.

Actually, from Equation \ref{eqn.rescaleddistribution}, we can
generalize an abstract form of the probability density function,
\begin{equation}\label{eqn.generalizedform}
f(x)\sim P^{-\beta} g(P^{-\beta} x).
\end{equation}
Where, $g(y)$ is an arbitrary probability density function with
size-independent argument $y$. We know that when we set $g(y)$ as
the concrete form, $\frac{(\alpha -1)^{\alpha}}{\Gamma
(\alpha)}\frac{\exp(-\frac{\alpha -1}{y})}{y^{1+\alpha}}$, we get
the generalized LV model in Equation \ref{eqn.rescaleddistribution}.

Actually, the power law relationship between population and GDP
which is discussed in Section \ref{sec.fromincometoallometry} can be
derived from the abstract form (Equation \ref{eqn.generalizedform})
because,
\begin{equation}\label{eqn.newmeanvalue}
\langle x\rangle = \int_0^{+\infty}{x f(x)dx}\sim \int_0^{+\infty}{x
P^{-\beta} g(P^{-\beta} x)dx},
\end{equation}
replace the integral variable $x$ with $y=P^{-\beta} x$, we obtain,
\begin{equation}\label{eqn.newmeanvalue1}
\langle x\rangle \sim \int_0^{+\infty}{P^{-\beta} P^{\beta} g(y)
P^{\beta}dy}=P^{\beta}\int_0^{+\infty}{g(y)dy}\sim P^{\beta},
\end{equation}
where, $\int_0^{+\infty}{g(y)dy}$ is a constant because $g(y)$ is
size-independent. Finally, we can also obtain the power law
relationship between population and GDP,
\begin{equation}\label{eqn.newallometry}
X=P \langle x\rangle\sim P^{1+\beta}.
\end{equation}
So, we can conclude that the essence of size-dependency in income
distribution is captured by Equation \ref{eqn.generalizedform}.
Actually, this re-scaled form distribution is not first discovered
by this paper. In
\citep{baldovin_central_2007,stella_anomalous_2010}, the authors
also gave a similar formula to describe the anomalous scaling
probability density function in critical phenomena,
\begin{equation}\label{eqn.limitdistribution}
f(x)\sim n^{-D} g(n^{-D} x).
\end{equation}
Where, $n$ is the size of the system (the number of addends), $D$ is
also a re-scaled exponent. The same mathematical form must imply the
ubiquitous natural laws, so the individual income can be viewed as a
sum of many correlated random variables related to each person in
the same country. However, we will not discuss the detail of this
discovery and leave it to the future studies because of the size
limitation of this paper.

\section{Concluding Remarks}
\label{sec.concluding}

This paper discussed the size-dependency of income distribution
which is a very important property and ignored more or less by
previous studies. The size-dependency has two important
implications: 1. The power law relationship between population and
GDP (which is also known as allometric growth); 2. The re-scaled
income distribution has the same mathematical form for the anomalous
scaling probability density function of the sum of many correlated
random variables. However, due to the limitation of our data, the
results discussed in this paper are only for United States of
America, this particular developed country, and only for the period
of 1996-2007 which is a very stable time of the United States. We
have observed that the allometric growth pattern is not found for
some countries, especially the nations encountering convulsions or
inflation by another data set. Thus, we hypothesize that the
size-dependency of income distribution, especially the power law
relationship between $\lambda$ and population will not be observed
as well in these cases.

In addition, we have found the same size-dependency phenomena in
human online behaviors \citep{wu_allometric_2010}, therefore, it is
reasonable to accept that some results in this paper as common ones
for the stable developing complex systems.

\begin{acknowledgments}
This paper is supported by National Natural Science Foundation of
China under Grant No. 61004107 and Grant No. 70771012.
\end{acknowledgments}

\bibliography{incometoallo}

\end{document}